\documentclass[twocolumn]{aastex701} 

\usepackage{multirow}

\usepackage{graphics,epsf}
\usepackage[utf8]{inputenc}
\usepackage{amsmath}                
\usepackage{amsfonts}               
\usepackage{amssymb}                
\usepackage{epsfig}                 
\usepackage{graphicx}               
\usepackage{float}
\usepackage{color}
\usepackage{multirow}               

\hypersetup{
    colorlinks=true,
    linkcolor=red,   
    urlcolor=cyan}

\usepackage[colorinlistoftodos]{todonotes}

\newcommand{\km}{{~\rm km}}
\newcommand{\s}{{~\rm s}}

\newcommand{\g}{{~\rm g}}
\newcommand{\G}{{~\rm G}}
\newcommand{\K}{{~\rm K}}
\newcommand{\erg}{{~\rm erg}}
\newcommand{\yr}{{~\rm yr}}

\newcommand{\pc}{{~\rm pc}}
\newcommand{\kpc}{{~\rm kpc}}

\newcommand{\mum}{{~\rm \mu m}}

\begin{document}
\title{Molecular bipolar outflow in SN 1987A supports the jittering-jets explosion mechanism}

\author[0000-0003-0375-8987]{Noam Soker}
\affiliation{Department of Physics, Technion - Israel Institute of Technology, Haifa, 3200003, Israel; soker@physics.technion.ac.il}
\email{soker@physics.technion.ac.il}

\begin{abstract}
I examine high-quality CO and SiO molecular maps of the core-collapse supernova (CCSN) remnant SN 1987A from the literature and find that the molecular gas exhibits a bipolar structure, correlated with the visible bipolar morphology (termed the keyhole) and the bipolar morphology of the iron emission map. I apply a method commonly used in the study of jet-shaped planetary nebulae. The keyhole has a morphology similar to that of many jet-shaped pairs of bubbles in cooling-flow clusters of galaxies and planetary nebulae. Therefore, the findings of this study, which make the bipolar structure of SN 1987A robust,  strengthen the claim that a pair of energetic jets shaped the keyhole and its surroundings. According to the jittering-jets explosion mechanism (JJEM), this pair of jets was the most energetic of several pairs that exploded SN 1987A. This study adds to the accumulating evidence that the JJEM is the primary explosion mechanism for CCSNe, responsible for the majority, if not all, CCSNe. 
\end{abstract}


\section{Introduction} 
\label{sec:intro}

The iconic core-collapse supernova (CCSN), now a CCSN remnant (CCSNR), SN 1987A, is one of the `battlegrounds' between the two competing theoretical explosion mechanisms that aim to explain CCSNe: the jittering-jet explosion mechanism (JJEM; \citealt{Soker2024UnivReview, Soker2025Learning} for recent reviews and \citealt{ShiranSoker2025, ShiranSoker2026, KlimovSoker2026, Soker2026J0450, Soker2026Failed, WangShishkinSoker2025} for recent papers) and the neutrino-driven mechanism (the delayed neutrino mechanism; e.g., \citealt{Janka2025, Janka2025Padova} for recent reviews and \citealt{Giudicietal2026, LuoZhaKajino2026, Murphyetal2026, Orlando2026, Rusakovetal2026, VarmaMuller2026} for some recent papers). Observables during the explosion, such as neutrino emission and the explosion energy of SN 1987A, cannot distinguish between the two theoretical explosion mechanisms (e.g., \citealt{Soker2025Learning}). Therefore, studies of the two explosion mechanisms compare the structure of SN 1987A ejecta with their predictions. 

Some studies have argued that the neutrino-driven mechanism can explain the basic structural properties of SN 1987A (e.g., \citealt{Jankaetal2017, Alpetal2019, Jerkstrandetal2020, Galberetal2021}). 
Some find that the neutrino-driven mechanism can explain some, but not all, properties of SN 1987A, e.g., \cite{McCrayFransson2016}, with references to \cite{Utrobinetal2015} and \cite{Wongwathanaratetal2015}.  
In \cite{Soker2017RAATwo}, I compared a three-dimensional (3D) iron structure of SN 1987A from \cite{Larssonetal2016} with 3D simulations of the neutrino-driven mechanism from \cite{Wongwathanaratetal2015}, and found that they disagree; I argued for the JJEM. 
Later papers within the JJEM framework that studied SN 1987A found that the JJEM can explain its large-scale morphology \citep{BearSoker2018, Soker2024NA1987A, Soker2024PNSN}. 
Most recently, in \cite{Soker2026Dust}, I referred to new observations of SN 1987A (e.g., \citealt{Bouchetetal2024, Matsuuraetal2024}) and strengthened the claim for a bipolar structure in relation to the dust distribution.   
Some more recent simulations also highlight the need for a bipolar explosion in SN 1987A (e.g., \citealt{Onoetal2020, Orlandoetal2020, Orlandoetal2025}). 

In a recent study, \cite{Wessonetal2026} compared the 3D structure of SN 1987A with 3D simulations within the framework of the neutrino-driven mechanism and argued for some similarities, although their simulations could not account for all of SN 1987A's morphological properties. I find the agreement between the simulations and the 3D structure of SN 1987A poor. To compare the SiO and CO ALMA observations with the simulations, they generated intensity maps using a minimum intensity threshold set to twice the noise threshold. In Section \ref{sec:Components}, I use their new observational images of the outer regions, as well as earlier ones, to identify outflow components in SN 1987A.  In Section \ref{sec:Bipolar}, I show that these outflow components fit the bipolar structure of SN 1987A, the 'keyhole'. In Section \ref{sec:Summary}, I argue that this structure is compatible with the JJEM.

\section{Outflow components} 
\label{sec:Components}

In what follows, I apply a method commonly used in the study of planetary nebulae, and which brought a deep and thorough understanding of the shaping of planetary nebulae by jets (e.g.,  \citealt{Morris1987, Soker1990AJ, Guerreroetal1998, Guerreoetal2021, Mirandaetal1998, SahaiTrauger1998, Sahaietal2005, Boffinetal2012, AkashiSoker2018, Sowickaetal2017, EstrellaTrujilloetal2019, RechyGarciaetal2019,  Tafoyaetal2019, Balicketal2020,   GarciaSeguraetal2020, GarciaSeguraetal2021, RechyGarciaetal2020, Clairmontetal2022, Danehkar2022, MoragaBaezetal2023, Ablimit2024, Derlopaetal2024, Mirandaetal2024, Sahaietal2024}). It begins with careful qualitative morphological identification and classification (e.g., \citealt{Balick1987, Parkeretal2006, Sahaietal2007, Kwok2024}), which nonetheless establishes robust identification of jet-shaped structures (e.g., \citealt{SahaiTrauger1998}). (For a recent quantitative analysis, see \citealt{ShishkinMichaelis2026}.) The qualitative and partially quantitative comparisons of simulations motivated by classification with observations (e.g., \citealt{GarciaSeguraetal2021, GarciaSeguraetal2022, GarciaSeguraetal2025}) have established, over the years, that jets are a major factor in shaping and powering outflows in planetary nebulae. Outflows from massive stars, which can become very complicated processes (e.g., \citealt{Michaelisetal2018, Gilkisetal2025, Grichener2025,  MukhijaKashi2025ApJ, MukhijaKashi2026ApJa, MukhijaKashi2026ApJb, MukhijaKashi2026NewA, MukhijaKashi2026PASP}), can also be shaped by jets (e.g., \citealt{AkashiKashi2020, Kashi2023}). The systematic comparison of 3D hydrodynamical simulations of CCSN jets in the framework of the JJEM is new (e.g., \citealt{Braudoetal2025, Braudoetal2026, SokerAkashi2025, AkashiSoker2026a, AkashiSoker2026BG11}). Therefore, any study, even if qualitative and short, that identifies jet-shaped structures in CCSNRs is of great value to the question of the explosion mechanism of CCSNe.

\subsection{Iron components} 
\label{sec:TwoAxes}

\begin{figure*}[]
	\begin{center}
\includegraphics[trim=0.0cm 12.9cm 0.0cm 0.0cm ,clip, scale=0.85]{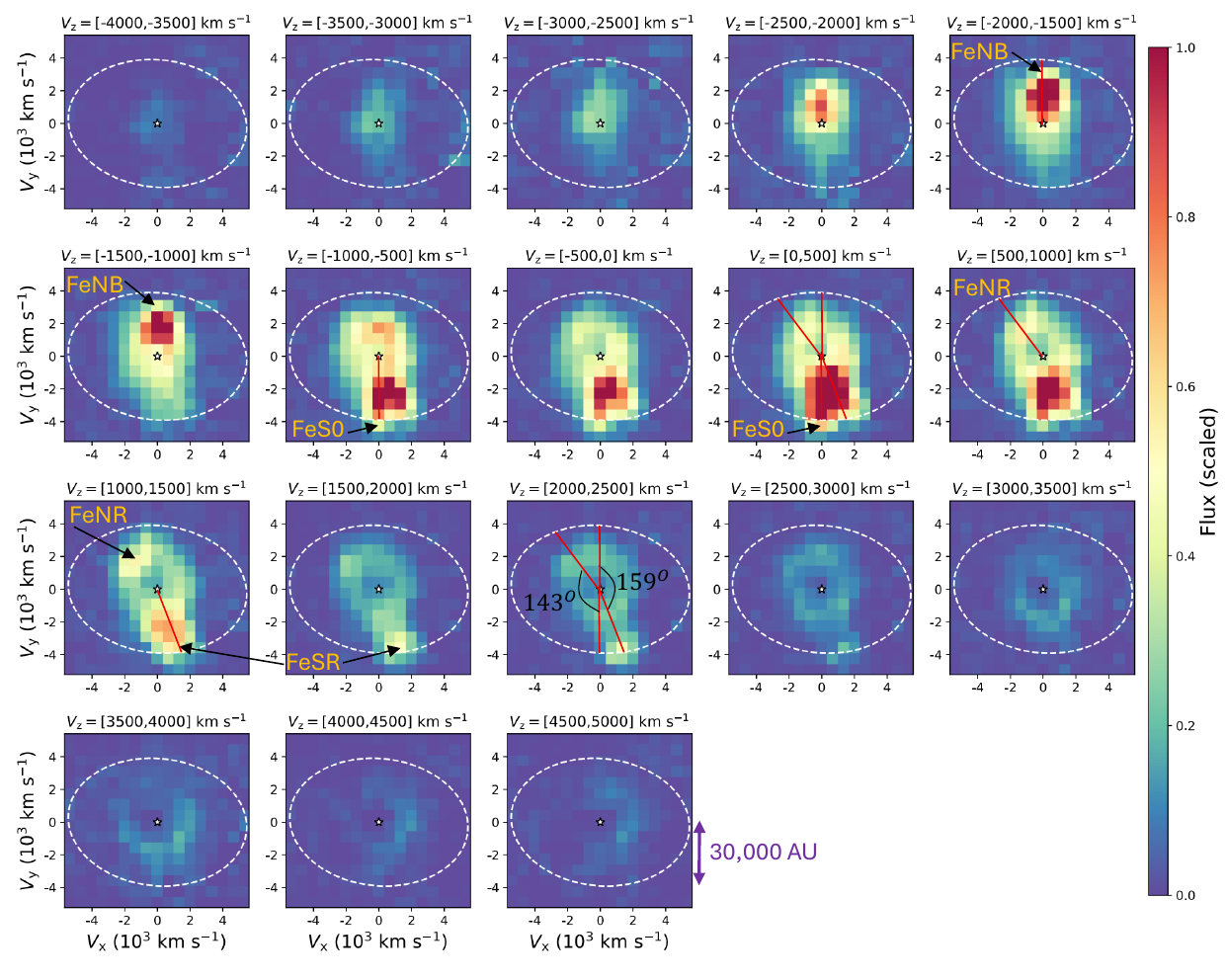} 
\caption{A figure adapted from \cite{Larssonetal2023}, presenting images of the [Fe \textsc{i}] $1.443 \mum$ emission line in Doppler velocity windows, as indicated. The dashed white line shows the position of the equatorial ring that the star lost $\simeq 20,000 \yr$ before the explosion. I added the red lines and the names of the outflow components: `Fe' indicates that the components are identified in the [Fe \textsc{i}] $1.443 \mum$ line, `N' and `S' mark north and south, respectively, and the last letter indicates whether the component is blueshifted, redshifted, or has about zero Doppler shift. The observation was at an age of 12,927 days after the explosion; one pixel in the image $0.1^{\prime \prime}=0.024 \pc$, corresponds to $664 \km \s^{-1}$ in the freely expanding ejecta, for the distance to the LMC of $49.6 \kpc$ that \cite{Larssonetal2023} assumed. }
\label{Fig:DopplerShift}
\end{center}
\end{figure*}

In this section, I identify outflow components that I relate to the bipolar structure of SN 1987A in Section \ref{sec:Bipolar}. Towards that goal, I start with a set of Doppler maps from \cite{Larssonetal2023} that I present in Figure \ref{Fig:DopplerShift}. I identify four bright, distinguishable outflow components and point to the tip of each. Two are bright regions, FeNB and FeS0, while two are much fainter, FeNR and FeSR; `Fe' indicates that the components are identified in the [Fe \textsc{i}] $1.443 \mum$ line, `N' and `S' mark north and south, respectively, and the last letter indicates whether the component is blueshifted, redshifted, or has about zero Doppler shift. 

 I identified the outflow components by visually inspecting the images. The center, common to all images, is from the original paper and is well defined. I then identified each of the four components by these criteria: ($i$) A pixel much brighter than its immediate surroundings (in half of the image) in at least one image. ($ii$) It appears as a bright pixel (even if not the brightest) in at least two other images. ($iii$) There are at least two other bright pixels adjacent to this pixel, in at least one image. 
These criteria robustly define the four components. The uncertainty in the direction is $\pm \Delta /2$, where $\Delta$ is the width of the pixel, at a distance of about $5 \Delta$, implying an uncertainty of about $\pm 6^\circ$ on the plane of the sky.

Component FeNB is a highly blueshifted component prominent in the $[-2000, -1000] \km \s^{-1}$ range. The red line indicates that this component points north (up) in these images. Its Doppler counterpart, FeSR, is the redshifted component and is prominent in the Doppler range $[1000, 2000] \km \s^{-1}$.  
FeS0 is the second bright region in the Doppler images, and it is prominent in the range of $[-1000, 1000] \km \s^{-1}$. The line that indicates this component points straight south (down). The component FeNR is prominent in the Doppler shift range of $[500, 1500] \km \s^{-1}$. On one panel in Figure \ref{Fig:DopplerShift}, I mark the angles between the lines. 

\subsection{Molecular maps} 
\label{sec:Maps}

\begin{figure*}[]
	\begin{center}
\includegraphics[trim=0.0cm 7.9cm 0.0cm 0.0cm ,clip, scale=0.85]{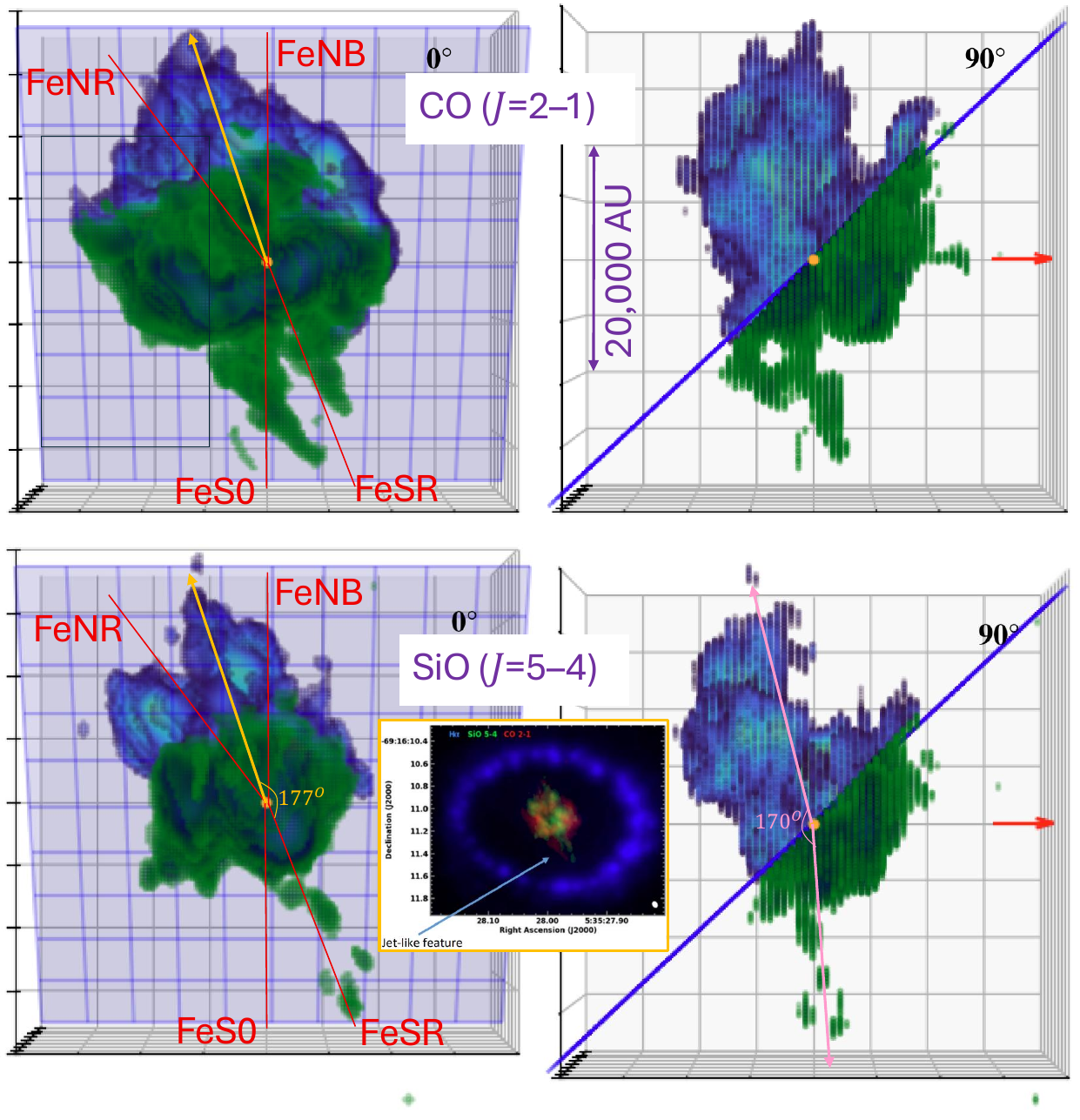} 
\caption{The four large panels are adapted from \cite{Wessonetal2026}. The two upper panels show 3D intensity maps of CO ($J=2-1$) emission, and the two lower panels show 3D intensity maps of SiO ($J=5-4$) emission. The left panel shows the plane of the sky; in the right panel, Earth is to the right (red arrows point toward Earth). The equatorial ring plane is indicated in blue. Blue regions are above the equatorial ring plane, and green material is below it. The orange point marks the explosion's position. Each side of a panel is 40,000 AU. 
I copied the four-line structures from Figure \ref{Fig:DopplerShift}, but the lengths of the lines here are $0.67$ times those in Figure \ref{Fig:DopplerShift} 
 so they fit in the images.   I also added the orange arrow in the left panels, which bisects the angle between outflow components FeNR and FeNB, and the two pink arrows in the lower-right panel to emphasize the bipolar structure of the molecular distribution.  
The inset between the two large lower panels is adapted from \cite{BearSoker2018} and \cite{Abellanetal2017}, showing CO ($J=2-1$) in red, SiO ($J=5-4$) in red, and H$\alpha$ in blue (i.e., the equatorial ring); the jet-like feature is marked from \cite{BearSoker2018}, who already identified the molecular south outflow. 
}
\label{Fig:Molecular1}
\end{center}
\end{figure*}

\cite{Wessonetal2026} construct 3D distributions of CO ($J=2-1$) and SiO ($J=5-4$) molecules from ALMA observations of SN 1987A. In Figure \ref{Fig:Molecular1}, I present their CO (upper panels) and SiO (lower panels) intensity maps on the plane of the sky (left panels) and with respect to an observer on the right (right panels), i.e., in a plane along the line of sight.  On the left panels, I overlaid the four-line structure that I built in Figure \ref{Fig:DopplerShift}, i.e., the lines in the direction of the four outflow components that I identified from the [Fe \textsc{i}] $1.443 \mum$ emission Doppler maps of \cite{Larssonetal2023}. The relative scale of the four lines is 0.67 times that in Figure \ref{Fig:DopplerShift}  so that they can fit into the images and serve for scaling.  I added the orange arrow in the left panels, which bisects the angle between outflow components FeNR and FeNB.   

Figure \ref{Fig:Molecular1} clearly shows that the molecular gas distribution has a large-scale bipolar structure that is correlated with the Fe outflow components. The bisector between outflow components FeNR and FeNB is along an elongated structure in CO and SiO towards the north-north-east. On the plane of the sky, but not in a plane along the line of sight, it is almost opposite to the component FeSR direction, along which there are clumps of SiO; the angle between the two is $177^\circ$. In the SiO map along the line of sight (lower-right panel of Figure \ref{Fig:Molecular1}), the SiO southern clumps are projected on the FeS0 component. In Section \ref{sec:Bipolar}, I suggest that the southern Fe outflow components and the far south molecular gas are concentrated around the nozzle, a faint radially extended zone in the keyhole (the optical bipolar bright structure) that I attributed to the action of one of the exploding jets \citep{Soker2024Keyhole}. The CO distribution correlated with the general direction of the FeS0 and FeSR components. The right panels of Figure \ref{Fig:Molecular1}, in the plane along the line of sight,  also reveal the bipolar structure of the molecular gas. I drew two pink arrows in the SiO map to emphasize this bipolar structure; the CO map also shows it. The two pink lines are $170^\circ$ apart.

 I drew the upper pink line from the center to the upper clump. The front of the blue region below the clump extends over $12^\circ$ as viewed from the center, $6^\circ$ to each side of the pink line. Therefore, I estimate the uncertainty in this direction to be about half this distance, or $\pm 3^\circ$. On the other side (lower region of the image), there are three clumps, so I drew the lower pink line through the central clump. The uncertainty in the direction is also $\pm 3^\circ$.

\cite{BearSoker2018} already hinted at this bipolar structure when referring to the south protrusion in a molecular map from \cite{Abellanetal2017} as a jet-like feature. On the plane of the sky, it is in the direction of the component FeSR. I present their image between the two lower panels of Figure \ref{Fig:Molecular1}. \cite{BearSoker2018} could not fully reveal the molecular gas bipolar structure. The new high-quality images from \cite{Wessonetal2026} enable me to clearly identify the molecular bipolar structure.  I will return to discuss this bipolar structure in Section \ref{sec:Bipolar}. 

\begin{figure}[]
	\begin{center}
\includegraphics[trim=0.0cm 3.0cm 0.0cm 0.0cm ,clip, scale=0.70]{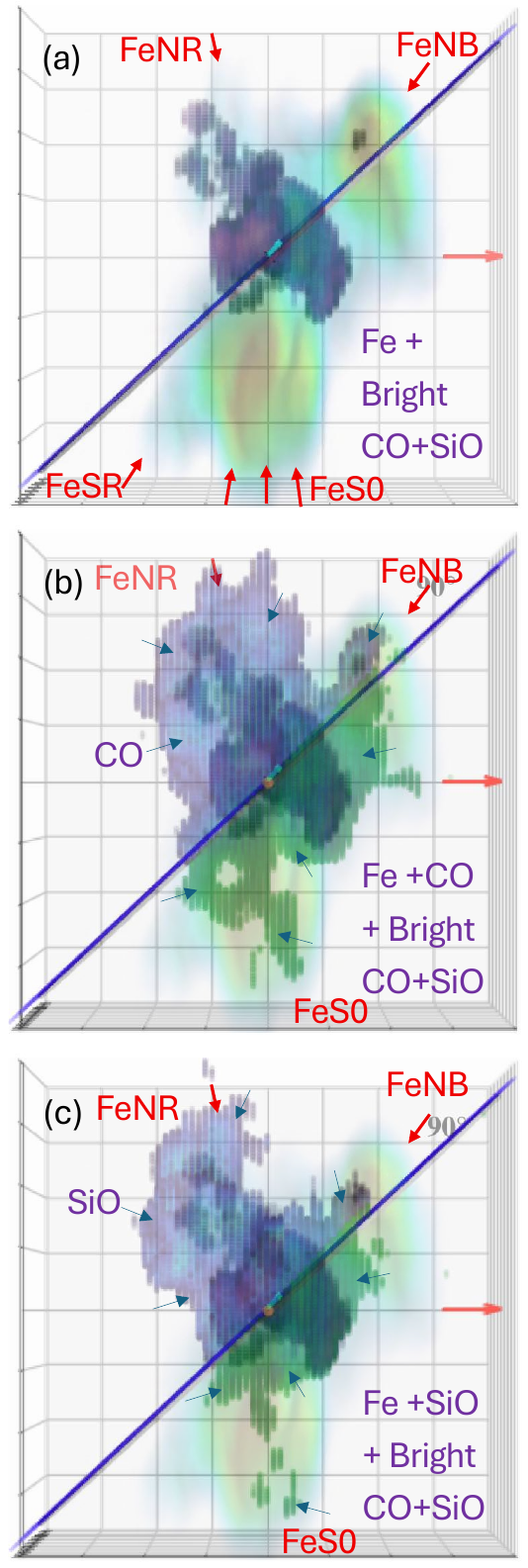} 
\caption{Images adapted from \cite{Wessonetal2026}, with my identification of the four outflow components. (a) Bright CO (dark purple-orange) and SiO (dark purple-green) emission overlaid with the [Fe \textsc{i}] (faint blue to yellow-red) emission that \cite{Wessonetal2026}
obtained from \cite{Larssonetal2023}. Only the regions with intensity above twice the noise threshold are shown by \cite{Wessonetal2026}. (b) A transparent full CO map from Figure \ref{Fig:Molecular1}, which I overlaid on the image from panel (a).  The full CO map contains the bright CO parts from panel (a) and extended regions beyond that; I point to 8 arbitrary locations of this fainter CO region with narrow blue arrows.  (c) A transparent full SiO map from Figure \ref{Fig:Molecular1}, which I overlaid on the image from panel (a).   The 8 narrow blue arrows point on the fainter SiO region. 
}
\label{Fig:Combined}
\end{center}
\end{figure}

\cite{Wessonetal2026} compare their observational findings with numerical simulations of explosions in the framework of the neutrino-driven mechanism. They argue that they can fit the observed CO morphology, but not the SiO velocity distribution and morphology. Specifically, their explosion simulations produce too little Si+O at high velocities, and overpredict the amount of low-velocity material. The JJEM can account for higher-velocity components with energetic jets  (Section \ref{sec:Summary}).   
The relevant point I make here is that to match the CO morphology, \cite{Wessonetal2026} examine only the brightest regions of the CO and SiO, i.e., those brighter than twice the noise threshold. In panel (a) of Figure \ref{Fig:Combined}, I present the image they compare to one of their simulations. By also presenting the faint regions of the molecular gas, I find the bipolar structure, which I further discuss in Section \ref{sec:Bipolar}.    

To demonstrate the correlation between the molecular morphology and the Fe outflow components, I overlaid a partially transparent intensity map of CO (panel b) and SiO (panel c) from the right panels of Figure \ref{Fig:Molecular1} on panel (a) of Figure \ref{Fig:Combined}. These panels show that the CO and SiO morphologies are correlated, at least in a line-of-sight plane, with the outflow components FeNR, FeNB, and FeS0. The extension of the molecular gas towards FeSR, which is in any case a faint component in [Fe \textsc{i}] $1.443 \mum$ emission, is marginal. 

This section shows that the outflow components defined by the [Fe \textsc{i}] $1.443 \mum$ emission and the molecular gas have a correlated bipolar outflow morphology. I turn to compare this with the visible band that exhibits the most prominent bipolar structure, the `keyhole.'
   
\section{Relation to the bipolar structure of SN 1987A} 
\label{sec:Bipolar}

\begin{figure}[]
	\begin{center}
\includegraphics[trim=0.0cm 15.0cm 0.0cm 0.0cm ,clip, scale=0.52]{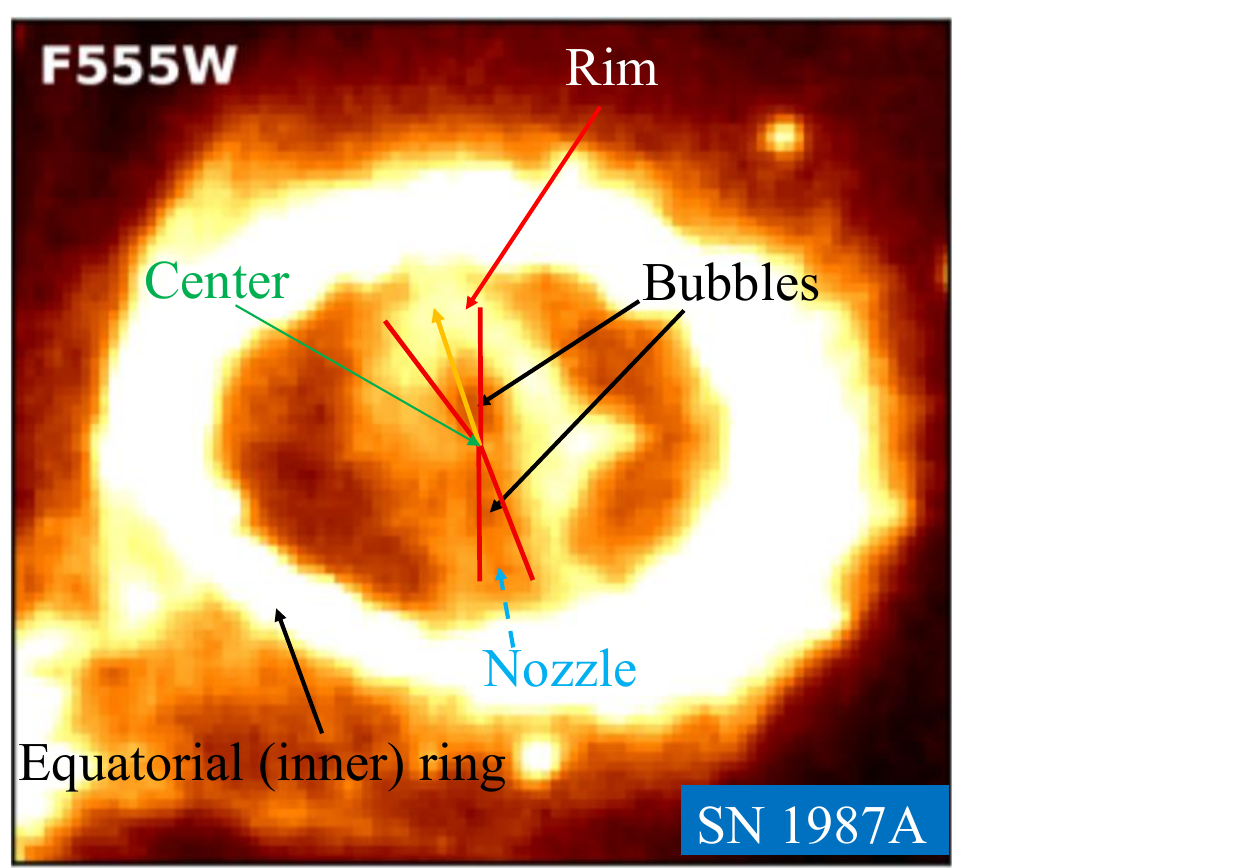} 
\caption{ A figure adapted from \cite{Soker2024Keyhole} which includes an HST/WFC3 image of SN 1987A in one filter from \cite{Rosuetal2024}; north is up and east to the left. The field of view is $2.50^{\prime \prime} \times 2.25^{\prime \prime}$. 
The `keyhole' is the north-south elongated bright structure within the equatorial ring. The identification of the rim, nozzle, and two bubbles is from \cite{Soker2024CFs} based on similarities to jet-inflated bubbles in cooling flow clusters of galaxies. 
Here, I added the structure of the four red lines and the orange arrow from Figure \ref{Fig:Molecular1}, at the same scale ($0.67$ times that of the four-line structure in Figure \ref{Fig:DopplerShift}).   
}
\label{Fig:Keyhole}
\end{center}
\end{figure}

HST and JWST observations reveal a bipolar structure of a northern low-intensity zone with a bright rim on its front and an elongated low-intensity nozzle in the south; this structure is termed `keyhole'.\footnote{See press release by JWST: \url{https://science.nasa.gov/missions/webb/webb-reveals-new-structures-within-iconic-supernova/}. }
The morphological similarity of some CCSNRs, including SN 1987A, with the bipolar morphologies of pairs of bubbles in cooling flow clusters of galaxies (e.g., \citealt{Soker2024CFs}) and planetary nebulae (e.g., \citealt{Soker2024PNSN}), brought me \citep{Soker2024CFs} to define the two bubbles and the rim-nozzle asymmetry in SN 1987A, as depicted in Figure \ref{Fig:Keyhole}, and to argue that jets shaped the bipolar structure of SN 1987A \citep{Soker2024Keyhole}. Specifically, that one pair of energetic jets was long-lived and shaped the keyhole (also \citealt{Soker20261987Afailed}). Here, I examine the relationship between molecular gas, Fe outflow components, and the keyhole's structure. 

On the keyhole image in Figure \ref{Fig:Keyhole}, I overlaid the structure of four red lines and the orange arrow that I constructed in Figure \ref{Fig:DopplerShift} and the left panels of Figure \ref{Fig:Molecular1}: The four red lines depict the Fe outflow components and the orange arrows the bisector of the two north components, which also coincide with the extended molecular gas structure to the north-north-east. In Figure \ref{Fig:Combined}, I showed that the molecular gas morphology coincides with three of the Fe outflow components. Figure \ref{Fig:Keyhole} shows that the bisector between the two northern outflow components is along the bubble and rim in the north, and the two southern outflow components are on the side of the nozzle in the south. 

The conclusion of this section is that the Fe outflow components and the extended structure of the molecular gas, namely, the highest velocity components of the molecular gas, are part of the bipolar structure. The southern Fe outflow components are around the nozzle, and the northern ones are around the axis through the center of the northern bubble. The most elongated part of the molecular gas, which is along the orange arrow to the north-north-east (Figure \ref{Fig:Molecular1}), is extended through the bubble and to the front of the rim (the orange arrow in Figure \ref{Fig:Keyhole}). As the earlier papers cited above (\citealt{Soker2024CFs, Soker2024PNSN}) mentioned, the bipolar rim-nozzle asymmetry with two bubbles is a common structure in cooling flow clusters and planetary nebulae, where it is attributed to shaping by jets.   
Most likely, a pair of energetic jets shaped the bipolar outflow of SN 1987A \citep{Soker2024Keyhole}. 

\section{Discussion and Summary} 
\label{sec:Summary}

In this short study, I examined the high-quality new analysis of ALMA observations of CO and SiO by \cite{Wessonetal2026}, which I present in Figures \ref{Fig:Molecular1} and \ref{Fig:Combined}. I find a correlation, although not perfect, between the molecular gas structure and the Fe outflow components that I identify in the Doppler maps by \cite{Larssonetal2023}, which I present in Figure \ref{Fig:DopplerShift}. Both the molecular gas structure and the Fe outflow components exhibit a general north-south bipolar structure. In Figure \ref{Fig:Keyhole}, I show that the bipolar structure of the [Fe \textsc{i}] $1.443 \mum$ emission outflow components, the fast regions of the molecular gas, and the optical image of SN 1987A, i.e., the keyhole, share the same general bipolar structure. The southern Fe outflow components and the molecular gas are concentrated around the nozzle.  I estimated (Section \ref{sec:Components}) the uncertainties in the directions of the different lines I drew to be $\lesssim 6^\circ$ (projected on the plane of the sky), implying that the bipolar morphology is robust.

 This study concentrated on the large-scale bipolar morphology. I argued that the Fe (Figure \ref{Fig:DopplerShift}), and the molecules CO and SiO (Figure \ref{Fig:Molecular1}) share the bipolar structure (Figure \ref{Fig:Combined}) of the keyhole (Figure \ref{Fig:Keyhole}). I also note, however, that their structures depend on differences in composition and excitation conditions. Reconstructing the ejecta three-dimensional structure also depends on assumptions such as homologous expansion, intensity threshold, and projection. Of these, the assumption of homologous expansion seems robust, as the keyhole did not yet interact with the inner ring of SN 1987A. For the above reasons, the apparent alignment does not automatically demonstrate that all three tracers belong to the same physical outflow. Nonetheless, the alignment I found in this study does suggest that these tracers share the bipolar structure.

As I discussed in earlier papers \citep{Soker2024CFs, Soker2024Keyhole}, the keyhole has a morphology similar to that of many jet-shaped pairs of bubbles in cooling-flow clusters of galaxies and planetary nebulae. 
In \cite{Soker2024PNSN}, I argued that the two opposite arcs outside the keyhole appearing in a JWST image by \cite{Matsuuraetal2024}, which share the same symmetrical axis as the keyhole, strengthen the claim for shaping by jets. Such arcs are common in jet-shaped bipolar planetary nebulae, and 3D numerical simulations of bipolar jets obtain them (e.g., \citealt{Akashietal2018}). 

 The JJEM simulations (e.g., \citealt{Braudoetal2025, Braudoetal2026, AkashiSoker2026a, AkashiSoker2026BG11, BraudoSoker2026}) are not yet at the stage to predict composition from nucleosynthesis or molecular structure in the ejecta, because these simulations at present follow only the first 10-20 seconds after the explosion. What these simulations do show is that along the jet direction (e.g., \citealt{Braudoetal2025}) material can move somewhat faster than the ejecta average expansion velocity. 
The molecules and dust form in the ejecta long after the explosion, when the jets are long gone, hence long after the jets dissipated their energy into the core of the exploding star (which becomes part of the ejecta). Also, the jets contain too little mass to explain the molecules and dust; these are formed in the ejecta. The jets can compress the ejecta and, by that, promote dust formation \citep{Soker2026Dust} and molecules.

This study made the bipolar structure more robust, e.g., by emphasizing the nozzle, thereby strengthening the claim that an energetic pair of jets shaped the bipolar structure of SN 1987A. According to the JJEM, this pair was not the only one to participate in the explosion of SN 1987A (e.g., \citealt{Soker20261987Afailed}), but was the most energetic.  Other pairs could form more opposite structural features, as I tentatively identified in SN 1987A \citep{Soker2024NA1987A}, i.e., a point-symmetric morphology. 
Point-symmetric morphologies of about twenty CCSNRs are the most challenging observation for the neutrino-driven mechanism. However, there are others (see reviews cited in Section \ref{sec:intro}). The next most challenging observational property is the high explosion energies of some CCSNe that are way above the reach of the energies that simulations of the neutrino-driven mechanism can obtain, $E_{\rm exp} \simeq 2 \times 10^{51} \erg$. For example, three out of six CCSNe studied by \cite{ZhaoJetal2026} are beyond the reach of the neutrino-driven mechanism. Adding a powerful magnetar will not help much, as such magnetars require a jet-driven explosion (e.g., \citealt{SokerGilkis2017, Kumar2025}).  

This study supports the claim that the JJEM is the primary explosion mechanism in CCSNe. Other energy sources can boost the jet-driven explosion (e.g., \citealt{Soker2026G11}), including neutrino heating \citep{Soker2022nu}, a magnetar (e.g., \citealt{SokerGilkis2017}), and, in rare cases, thermonuclear burning triggered by the collapse of a pre-collapse mixed layer of helium and oxygen (\citealt{Klimovetal2026};  for the thermonuclear process itself see, e.g., \citealt{KushnirKatz2015, BlumKushnir2016}).

\section*{Acknowledgments} 
 I thank anonymous referees for helpful comments. 
A grant from the Pazy Foundation 2026 supported this research.
I thank the Charles Wolfson Academic Chair at the Technion. 





 \bibliography{reference}{}
  \bibliographystyle{aasjournal}
 
\end{document}